\documentclass[sigconf]{acmart}
\usepackage{hyperref}
\usepackage{url}
\usepackage{graphicx}

\usepackage{enumitem}

\usepackage{multirow}

\usepackage[table]{xcolor}
\AtBeginDocument{%
  }

\usepackage[most]{tcolorbox}

\newtcolorbox{promptbox}[1]{
    colback=gray!5!white,       
    colframe=gray!75!black,     
    fonttitle=\bfseries,        
    title=#1,                   
    arc=2pt,                    
    outer arc=2pt,
    left=5pt, right=5pt,        
    boxrule=0.5pt,              
    enhanced,
    breakable,                  
}

\setcopyright{acmlicensed}
\copyrightyear{2018}
\acmYear{2018}
\acmDOI{XXXXXXX.XXXXXXX}
\acmConference[Conference acronym 'XX]{Make sure to enter the correct
  conference title from your rights confirmation email}{June 03--05,
  2018}{Woodstock, NY}
\acmISBN{978-1-4503-XXXX-X/2018/06}

\begin{document}

\title{Generative Pseudo-Labeling for Pre-Ranking with LLMs}

\author{Junyu Bi}
\authornote{Both authors contributed equally to this research.}
\affiliation{%
  \institution{Alibaba Group}
  \city{Beijing}
  \country{China}
}
\email{bijunyu.bjy@taobao.com}

\author{Xinting Niu}
\authornotemark[1]
\affiliation{%
  \institution{Alibaba Group}
  \city{Beijing}
  \country{China}
}
\email{niuxinting.nxt@alibaba-inc.com}

\author{Daixuan Cheng}
\authornote{Work done during an internship at Alibaba Group.}
\affiliation{%
  \institution{Renmin University}
  \city{Beijing}
  \country{China}}
\email{daixuancheng6@gmail.com}

\author{Kun Yuan}
\authornote{Corresponding author.}
\affiliation{%
  \institution{Alibaba Group}
  \city{Beijing}
  \country{China}}
\email{yuankun.yuan@taobao.com}

\author{Tao Wang}
\affiliation{%
  \institution{Alibaba Group}
  \city{Beijing}
  \country{China}}
\email{wt439443@taobao.com}

\author{Binbin Cao}
\affiliation{%
  \institution{Alibaba Group}
  \city{Beijing}
  \country{China}}
\email{simon.cbb@taobao.com}

\author{Jian Wu}
\affiliation{%
  \institution{Alibaba Group}
  \city{Beijing}
  \country{China}}
\email{joshuawu.wujian@taobao.com}

\renewcommand{\shortauthors}{Junyu Bi et al.}

\begin{abstract}
Pre-ranking is a critical stage in industrial recommendation systems, tasked with efficiently scoring thousands of recalled items for downstream ranking. A key challenge is the train–serving discrepancy: pre-ranking models are trained only on exposed interactions, yet must score all recalled candidates—including unexposed items—during online serving. This mismatch not only induces severe sample selection bias but also degrades generalization, especially for long-tail content. Existing debiasing approaches typically rely on heuristics (e.g., negative sampling) or distillation from biased rankers, which either mislabel plausible unexposed items as negatives or propagate exposure bias into pseudo-labels. In this work, we propose \textbf{Generative Pseudo-Labeling} (GPL), a framework that leverages large language models (LLMs) to generate unbiased, content-aware pseudo-labels for unexposed items, explicitly aligning the training distribution with the online serving space. By offline generating user-specific interest anchors and matching them with candidates in a frozen semantic space, GPL provides high-quality supervision without adding online latency. Deployed in a large-scale production system, GPL improves click-through rate by 3.07\%, while significantly enhancing recommendation diversity and long-tail item discovery.

\end{abstract}




\maketitle
\section{Introduction}

\begin{figure}[t]
\centering
\includegraphics[width=0.45\textwidth]{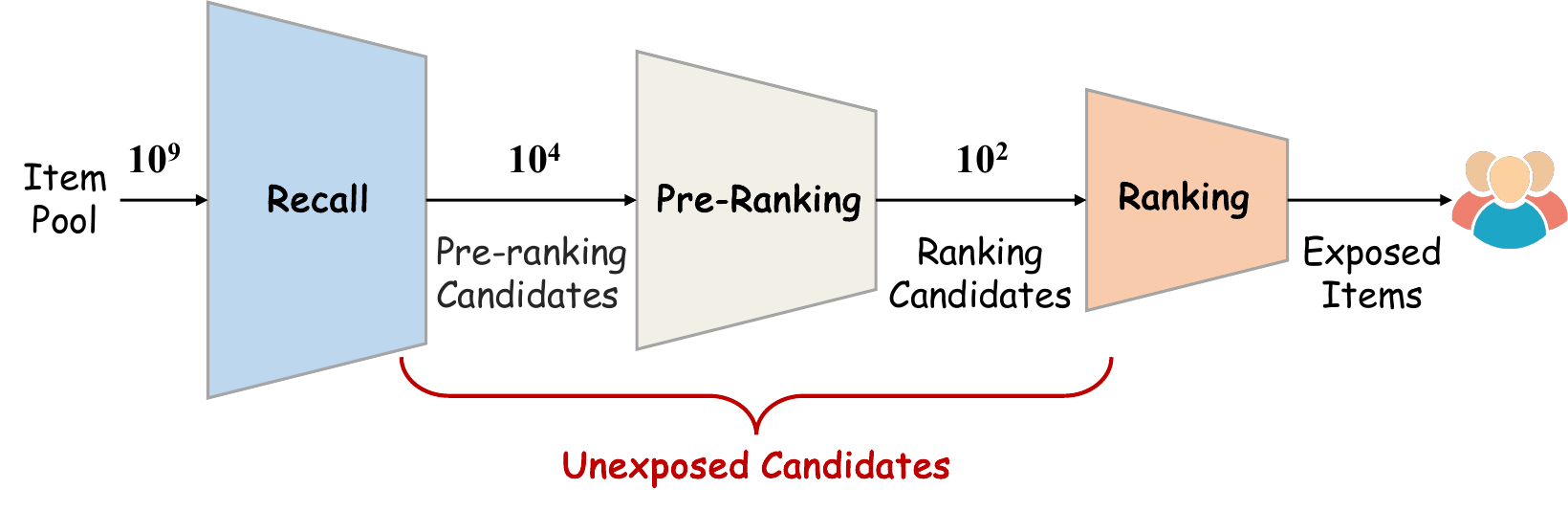}
\caption{Illustration of the cascade-based architecture in recommender systems. The unexposed candidates are items retrieved in early stages but discarded.}
\label{intro}
\vspace{-4pt}
\end{figure}

In large-scale industrial recommendation systems, cascade architectures—comprising candidate recall, pre-ranking, ranking, and re-ranking—are widely used~\citep{zhang2019deep} (Fig.~\ref{intro}). The pre-ranking stage scores thousands of recalled items with lightweight models, forwarding only the top candidates to ranking. However, this early filtering exacerbates sample selection bias (SSB)~\citep{ovaisi2020correcting}: during training, pre-ranking models rely solely on exposed interactions (e.g., clicks or purchases), yet at inference must score all recalled items—including unexposed ones lacking feedback. This train–inference distribution shift harms generalization, especially for long-tail or novel items, promoting over-recommendation of popular content, reinforcing information cocoons, and reducing diversity~\citep{xi2025bursting}.

Research on mitigating SSB has pursued several directions: negative sampling introduces false negatives~\citep{qin2022rankflow}; randomized exploration is costly~\citep{liu2023bounding}; teacher–student distillation inherits exposure bias~\citep{gao2023rec4ad}; and adversarial alignment is unstable~\citep{xing2021algorithmic}.
Meanwhile, LLMs have demonstrated strong semantic priors in recommendation~\citep{zhang2025llm}, yet no existing work leverages them to infer relevance for unexposed items—leaving a critical gap in addressing SSB during pre-ranking.

To bridge the train–inference gap and mitigate exposure bias, we propose \textbf{Generative Pseudo-Labeling} (GPL), a two-stage framework that aligns pre-ranker training with the full recall set encountered at serving time.
\textbf{Firstly}, GPL generates content-aware pseudo-labels for unexposed items. It begins by tokenizing items into hierarchical \emph{semantic identifiers} (SIDs) using a frozen multimodal encoder and a Residual Quantized VAE—ensuring representations are grounded in content, not biased interactions. A fine-tuned large language model then predicts future SIDs from user histories, which are decoded into \emph{interest-anchors}. Relevance scores are computed by matching these anchors with candidates in the frozen multimodal space, and further calibrated via an uncertainty-aware weighting scheme that accounts for semantic coherence, historical consistency, and LLM confidence.
\textbf{Secondly}, the pre-ranker is trained on a unified objective that jointly optimizes observed interactions and the calibrated pseudo-labels through a confidence-weighted loss. Crucially, all LLM inference and pseudo-label generation are performed offline and cached per user, imposing zero latency overhead during online serving.
Our contributions are as follows:

\begin{itemize}[leftmargin=*]
\itemsep0em
\item We propose GPL, to the best of our knowledge the first framework that leverages LLMs to address SSB in pre-ranking by explicitly aligning the training supervision with the online serving space, enabling robust generalization to unexposed and long-tail items. GPL overcomes the limitation of relying solely on exposed data, alleviates SSB, and enables the modeling of a broader and more diverse spectrum of user interests.

\item We design an efficient two‑step pseudo‑labeling process for unexposed items, which on the one hand reduces reliance on exposed samples through generative interest‑anchors creation, and on the other hand leverages multimodal information to further mitigate exposure and popularity biases. 

\item The method has been deployed on a large‑scale industrial recommendation platform, where it achieved a 3.07\% increase in click‑through rate (CTR) in online A/B testing, and simultaneously enhancing recommendation diversity while alleviating long‑tail distribution bias.
\end{itemize}

\section{Related Work}
In industrial recommender systems, the pre-ranking stage quickly filters large candidate sets but often suffers from SSB due to training exclusively on exposed interactions. Meanwhile, LLMs have recently demonstrated strong potential in recommendation via semantic understanding and knowledge reasoning. We thus review related work along two directions: (1) SSB mitigation methods in recommender systems, and (2) LLM applications in recommendation.

\subsection{Mitigating SSB in Recommender Systems}
A first family of approaches reshapes training data. Binary classification or negative sampling treats unexposed items as negatives~\citep{chen2023bias, zhao2025hybrid}, which inevitably introduces false negatives and reinforces popularity bias. Randomized content delivery collects unbiased feedback but is costly and impractical at scale~\citep{liu2023bounding, gao2023rec4ad}.

To bypass exploration, teacher–student transfer distills ranker scores as soft pseudo-labels~\citep{song2025knowledge, gao2025both}, but the teacher's exposure bias propagates into pseudo-labels. Adversarial domain adaptation~\citep{li2025unbiased, lin2024mitigating} aligns exposed and unexposed distributions, yet remains unstable and dependent on biased interaction graphs~\citep{xing2021algorithmic}.

\subsection{Applications of LLMs for Recommendation}
LLMs introduce rich semantic priors and weak supervision derived from item content, enabling diverse recommendation tasks: \emph{metadata enrichment}~\citep{hu2024enhancing}, \emph{intent elicitation}~\citep{lian2025egrec}, \emph{candidate generation}~\citep{du2024enhancing}, and \emph{synthetic feedback}~\citep{zhang2025llm}—all less prone to exposure bias than behavioral signals.

However, none of these works specifically targets \emph{unexposed items}. Our work fills this gap by fusing LLM-generated semantic patterns with observed interactions via confidence-aware calibration, providing a plug-and-play solution to mitigate SSB without expensive unbiased logging or joint end-to-end retraining.

\section{Methodology}
\subsection{Problem Formulation}
We consider a recommendation setting with a set of users $\mathcal{U}$ and an item corpus $\mathcal{H}$. The goal is to learn a relevance scoring function $f: \mathcal{U} \times \mathcal{H} \rightarrow \mathbb{R}$ that ranks items for each user, with higher scores indicating greater interest.
In practice, supervision comes only from \emph{exposed interactions}:
\[
\mathcal{D}_{\mathrm{e}} = \{(u, h, y_{u,h}) \mid u \in \mathcal{U},\, h \in S_u \subseteq \mathcal{H},\, y_{u,h} \in \{0,1\} \},
\]
where $S_u$ is the set of items shown to user $u$, and $y_{u,h}$ denotes binary feedback (e.g., click). However, exposure is biased toward popular items, limiting the model’s ability to discover long-tail content.

To address this, we additionally leverage \emph{unexposed candidates}:
\[
\mathcal{D}_{\mathrm{u}} = \{(u, h) \mid u \in \mathcal{U},\, h \in \overline{S}_u \subseteq \mathcal{H} \setminus S_u \},
\]
which consists of items retrieved by upstream stages (e.g., recall) but not exposed to the user. Though lacking ground-truth labels, these pairs are semantically plausible and offer opportunities for discovery.
The core challenge is to assign reliable pseudo-labels $r_{u,h} \in [0,1]$ to $(u,h) \in \mathcal{D}_{\mathrm{u}}$. We tackle this via a content-aware preference estimator (Section~\ref{sec:alignment}) and train $f$ with a joint objective:
\begin{equation}
    \min_f \; \mathcal{L}_{\text{al}}(\mathcal{D}_{\mathrm{e}}) + \lambda \, \mathcal{L}_{\text{pl}}(\mathcal{D}_{\mathrm{u}}),
    \label{eq:objective}
\end{equation}
where $\mathcal{L}_{\text{al}}$ and $\mathcal{L}_{\text{pl}}$ are losses on actual and pseudo labels, respectively, and $\lambda > 0$ balances their contributions.

\subsection{Overview of the Framework}
GPL addresses exposure bias in two stages. First, items are tokenized into content-based semantic identifiers (SIDs) via a multimodal RQ-VAE, and a fine-tuned LLM predicts future SIDs from user histories to produce interest anchors. These anchors are matched with unexposed candidates in a shared multimodal space to generate uncertainty-calibrated pseudo-labels. Second, the pre-ranker is jointly optimized on observed interactions and pseudo-labeled unexposed items.

\begin{figure*}[t]
\centering
\includegraphics[width=1.0\textwidth]{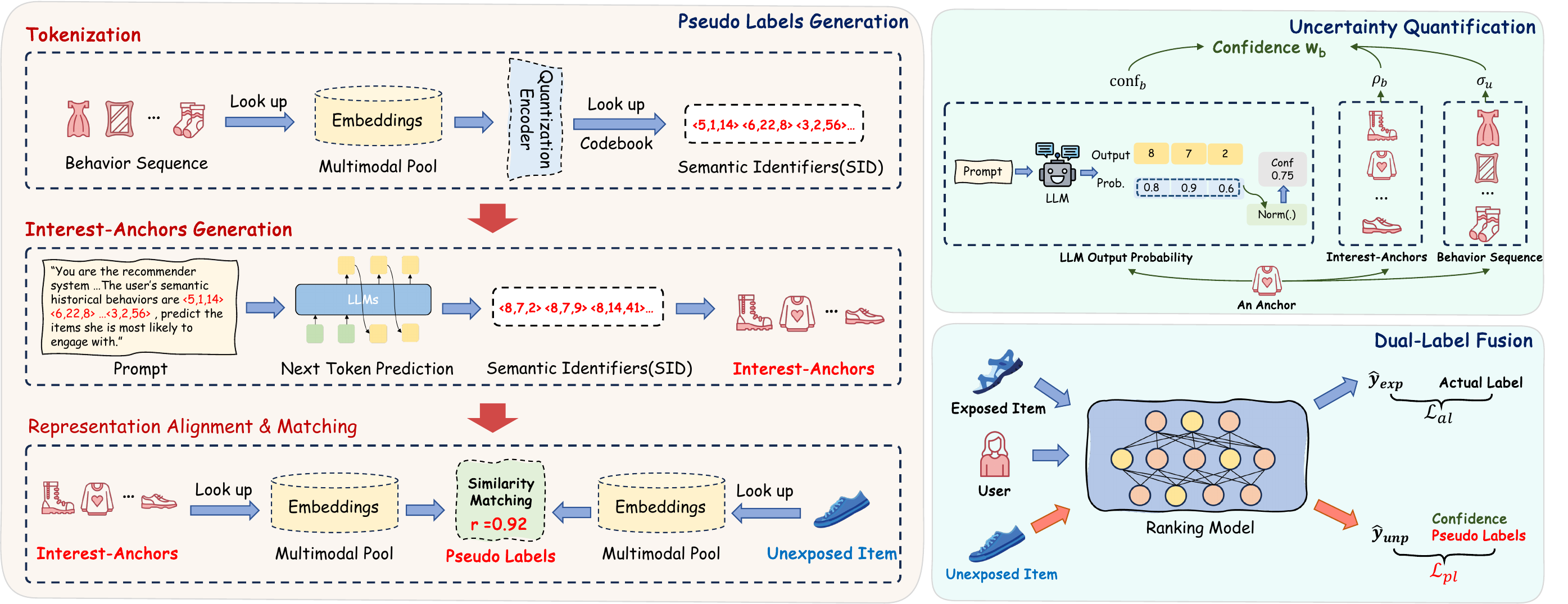}
\vspace{-6pt}
\caption{Overview of the GPL framework. For each user, interest anchors are generated offline via an LLM in a discrete semantic space and matched with unexposed candidates to produce pseudo-labels. These are calibrated by confidence weights based on semantic coherence, historical consistency, and generation uncertainty. In the Dual-Label Fusion stage, pseudo-labeled unexposed items are combined with exposed items bearing actual labels to jointly train the pre-ranking model.}
\label{model}
\vspace{-4pt}
\end{figure*}

\subsection{Pseudo Labels Generation}

\subsubsection{\textbf{Tokenization of User Behaviors}}
To enable symbolic reasoning with LLMs while avoiding exposure bias, we tokenize items into discrete semantic identifiers (SIDs) using only content information.
Each item \( h \) is first encoded by a frozen pre-trained multimodal encoder \( f_e \), which fuses textual descriptions and visual features into a unified embedding \( \mathbf{e}_{\text{mm}} = f_e(h) \in \mathbb{R}^d \). These embeddings are precomputed for all items and stored in a shared multimodal embedding pool, allowing any module in the pipeline to retrieve an item’s representation via efficient lookup. The encoder remains frozen to preserve general-purpose semantics and prevent contamination from biased interactions.

We train a Residual Quantized Variational Autoencoder (RQ-VAE)~\citep{van2017neural, sun2020vqrec} on the full item corpus, using \( \mathbf{e}_{\text{mm}} \) as reconstruction targets. The RQ-VAE learns \( L \) hierarchical codebooks that quantize \( \mathbf{e}_{\text{mm}} \) into a sequence of discrete indices \( s = (s^{(1)}, \dots, s^{(L)}) \), defined as the item’s SID. Critically, training relies solely on reconstruction and commitment losses, ensuring the token space reflects intrinsic content rather than exposure patterns.
During training, we build a lookup table mapping each observed SID to its item ID, enabling exact resolution at inference. User interaction histories are converted into SID sequences for LLM processing. This content-only tokenizer yields a semantically grounded, content-decoupled discrete space—particularly suitable for long-tail and cold-start scenarios.

\subsubsection{\textbf{Interest-Anchors Generation}} 
Given the tokenized user history $H_u$, we adapt a pre-trained open-source LLM to predict future interests in the semantic identifier (SID) space. The model input is constructed as:
\begin{equation}
    X_u = \mathrm{concat}\big( \mathrm{tokens}(T),\ \mathrm{tokens}(H_u) \big),
\end{equation}
where $T$ is a natural-language instruction prompt (details in our online supplement). 
The LLM is trained exclusively on sequences of positively interacted items (e.g., clicks or purchases), converted to their SIDs. To mitigate popularity bias inherited from interaction logs, we apply frequency-based downsampling to the top 10\% most frequent items in the training corpus. Each interaction involving such an item is retained with probability inversely proportional to its frequency~\citep{zhang2021causal}. This strategy reduces the dominance of head items while preserving sufficient signal for stable LLM training. 

To balance adaptation efficacy and computational efficiency, we employ Low-Rank Adaptation (LoRA)~\citep{hu2022lora}(rank=8): trainable low-rank matrices are injected into the query and value projections of all attention layers, while the original LLM parameters remain frozen. This reduces trainable parameters by over 99\% and preserves the model’s pre-trained semantic knowledge. 
Training follows the Next Token Prediction (NTP) objective. The last SID in $H_u$ is masked as the target, and the loss minimizes the negative log-likelihood over all hierarchical levels:
\begin{equation}
    \mathcal{L}_{\text{ntp}} = \sum_{l=1}^{L} -\log P\left( s_{\text{next}}^{(l)} \,\big|\, X_u, s_{\text{next}}^{(1:l-1)}; \theta \right),
\end{equation}
where $\theta$ denotes the trainable parameters.

At inference, we perform hierarchical beam search~\cite{beamsearch} with width $B$ to decode diverse SID candidates $\mathcal{S}_{\text{beam}} = \{ \mathbf{s}_1, \dots, \mathbf{s}_B \}$. Each generated SID $\mathbf{s}_b$ is resolved to a real item $a_b$ via the precomputed lookup table built during RQ-VAE training. Empirically, over 98.5\% of generated SIDs exist in the table (measured on a validation cohort), confirming strong coverage of plausible future interactions. For the rare out-of-vocabulary cases (<1.5\%), we fall back to nearest-neighbor retrieval in the frozen multimodal embedding space. The final interest-anchor set is $A_u = \{ a_1, \dots, a_B \}$. Throughout training and inference, a causal attention mask enforces strict temporal order, preventing information leakage and better capturing sequential user dynamics.

\subsubsection{\textbf{Representation Alignment \& Matching}} \label{sec:alignment}
We employ a large pre-trained multimodal encoder \( f_e \) to project both interest anchors and unexposed items into a shared semantic space. The encoder fuses heterogeneous inputs—such as product images, titles, price, category, and metadata—to produce rich, content-aware embeddings. By aligning user interests and candidates in this semantically coherent space, relevance scoring relies on content-based matching rather than interaction history, explicitly mitigating exposure and popularity biases and enhancing discoverability of long-tail items with rich content but sparse interactions.

Given these aligned representations, the relevance score for a user–item pair \((u, h)\) is computed as follows. For each anchor \( a_i \in A_u \), we compute cosine similarity \(\mathrm{cos}(f_e(a_i), f_e(h))\). While a naive approach might average over all anchors, uniform averaging assumes equal relevance—an assumption often violated when \(A_u\) contains noisy or weakly related candidates due to LLM generation errors or embedding distortions~\citep{li2019collaborative,zhang2023robust}. Prior work in multi-interest modeling shows that max-pooling is more robust, as it emphasizes the strongest semantic match while suppressing irrelevant signals~\citep{li2019multi}. Following this principle, we define:
\begin{equation}
\label{conf_w}
    r_{u,h} = Sigmoid(\frac{\max_{a_i \in A_u} \mathrm{cos}(f_e(a_i), f_e(h))}{\tau}) \in (0, 1),
\end{equation}
where $\tau > 0$ is a temperature hyperparameter tuned on a validation cohort to optimize ranking calibration. This score serves as the pseudo-label for \((u, h)\) in the subsequent \emph{dual-label fusion} stage.

\subsection{Dual-Label Fusion}
\subsubsection{\textbf{Uncertainty Quantification}}
To assess the reliability of LLM-generated interest anchors, we quantify uncertainty along three orthogonal dimensions:

\begin{itemize}[left=0pt]
    \item \textbf{Semantic Dispersion}: 
    Given the $B$ anchor embeddings $\{f_e(a_b)\}_{b=1}^B$, we compute their pairwise dissimilarity:
    \begin{equation}
        \sigma_u = \frac{1}{B(B-1)} \sum_{i \neq j} \big(1 - \cos(f_e(a_i), f_e(a_j))\big).
    \end{equation}
    A high $\sigma_u$ indicates unstable generation and reflects epistemic uncertainty.

    \item \textbf{Historical Consistency}: 
    For each anchor $a_b$, we measure its alignment with the user’s historical interactions $\{h_m\}_{m=1}^M$:
    \begin{equation}
        \rho_b = \frac{1}{M} \sum_{m=1}^M \cos\big(f_e(a_b), f_e(h_m)\big).
    \end{equation}

    \item \textbf{LLM-Intrinsic Confidence}: 
    For anchor $a_b$ corresponding to SID $\mathbf{s}_b$, we compute its average log-probability along the decoding path~\cite{gpt}:
    \begin{equation}
        \text{conf}_b = \frac{1}{L} \sum_{l=1}^L \log P\big(s_b^{(l)} \,\big|\, X_u, s_b^{(1:l-1)}; \theta \big).
    \end{equation}
\end{itemize}
These three uncertainty signals capture complementary aspects of anchor reliability. \textbf{Semantic Dispersion} ($\sigma_u$) reflects the \emph{epistemic uncertainty} of the LLM: if generated anchors are semantically inconsistent, the model is likely uncertain about the user’s intent. \textbf{Historical Consistency} ($\rho$) measures \emph{behavioral plausibility}: an anchor that aligns well with the user’s past interactions is more likely to represent a genuine preference. \textbf{LLM-Intrinsic Confidence} ($\text{conf}$) captures the \emph{aleatoric uncertainty} from the generation process itself—low log-probability indicates the LLM is extrapolating beyond its training distribution. By jointly considering these dimensions, our weighting scheme down-weights anchors that are either internally inconsistent, behaviorally implausible, or generated with low confidence, thereby improving the quality of pseudo-labels.

We first calculate the anchor-level confidence weight for anchor $a_b$ as:
\begin{equation}
    \label{eq:anchor_level}
    \tilde{w}_b = \exp(\lambda_1 \rho_b) \cdot \exp(\lambda_2 \cdot \text{conf}_b),
\end{equation}
and then apply user-level calibration based on semantic dispersion $\sigma_u$, which quantifies the coherence of generated anchors. The final confidence weight is obtained via softmax normalization:
\begin{equation}
    w_b = \exp(-\sigma_u) \cdot \frac{\exp(\tilde{w}_b)}{\sum_{b'=1}^B \exp(\tilde{w}_{b'})},
    \label{lab:conf}
\end{equation}
where the term $\exp(-\sigma_u)$ uniformly down-weights all anchors for users with high semantic dispersion, reflecting reduced overall reliability.

\subsubsection{\textbf{Joint Optimization}}
We train the lightweight pre-ranker $f$ with a unified probabilistic objective. For exposed data $\mathcal{D}_{\mathrm{e}}$, we use standard binary cross-entropy (BCE) with hard labels $y_{u,h} \in \{0,1\}$:
\begin{equation}
\mathcal{L}_{\text{al}} = -\frac{1}{|\mathcal{D}_{e}|} 
\sum_{(u,h) \in \mathcal{D}_{e}} \left[ y_{u,h} \log \hat{y}_{u,h} + (1 - y_{u,h}) \log (1 - \hat{y}_{u,h}) \right],
\end{equation}
where $\hat{y}_{u,h} = f(u, h)$.

For unexposed data $\mathcal{D}_{\mathrm{u}}$, we first calculate the confidence weight for each pair $(u, h)$: let $a^* = \arg\max_{a_i \in A_u} cos(f_e(a_i), f_e(h))$ be the anchor yielding the strongest match. The sample-level confidence weight is then $w_{u,h} = w_{a^*}$, as defined in Eq~(\ref{lab:conf}).
The pseudo-label loss is then defined as a confidence-weighted BCE:
\begin{equation}
\mathcal{L}_{\text{pl}} = -\frac{1}{|\mathcal{D}_{u}|} 
\sum_{(u,h) \in \mathcal{D}_{u}} w_{u,h} \left[ r_{u,h} \log \hat{y}_{u,h} + (1 - r_{u,h}) \log (1 - \hat{y}_{u,h}) \right].
\end{equation}

The final training objective combines both actual labels and pseudo labels in a unified manner:
\begin{equation}
\mathcal{L} = \mathcal{L}_{\text{al}} + \lambda \mathcal{L}_{\text{pl}},
\end{equation}
where $\lambda$ balances the contribution of pseudo-labels. In practice, we select $\lambda$ and $\tau$ jointly via online A/B testing to maximize click-through rate (CTR) and long-tail coverage. All components except the pre-ranker $f$ remain frozen during training.

\subsection{Industrial Deployment Details}
The pseudo-label pipeline runs entirely offline, imposing zero runtime overhead. Multimodal embeddings are precomputed at system initialization, L2-normalized, and cached in a distributed feature store; the encoder is re-invoked only for newly uploaded items, while existing embeddings are reused. The only large-model computation is LLM-based interest-anchor generation: by operating at the user level (one forward pass per user) rather than per candidate, inference calls are reduced by $\sim$6$\times$. On 96 NVIDIA H20 GPUs, one day's user traffic completes within six hours. Subsequent steps (lookup, similarity matching, confidence weighting) run on standard CPUs, and the resulting pseudo-labels are merged with observed interactions to train the pre-ranker—requiring no modification to the online serving architecture.

\begin{table*}[t]
    \caption{Overall performance comparison on the Taobao industrial dataset and Taobao-MM dataset.}
    \vspace{-8pt}
    \centering
    \scalebox{0.95}{
    \setlength{\tabcolsep}{1.5mm}
    {\begin{tabular}{l|ccc|c|c|ccc|c|c}
\toprule
Dataset & \multicolumn{5}{c|}{\textbf{Industrial Dataset}} &  \multicolumn{5}{c}{\textbf{Taobao-MM}} \\
\midrule
Method & HR@3 & HR@5 & HR@10 & AUC & GAUC & HR@3 & HR@5 & HR@10 & AUC & GAUC\\
\midrule
BC & 0.4863 & 0.5850 & 0.7611 & 0.7096 & 0.6098 & 0.4412 & 0.5405 & 0.7122 & 0.6854 & 0.5912 \\
KD & 0.4938 & 0.6247 & 0.8181 & 0.7110 & 0.6114 & 0.4498 & 0.5721 & 0.7516 & 0.6892 & 0.5945 \\
TL & 0.4905 & 0.6283 & 0.8102 & 0.7106 & 0.6107 & 0.4470 & 0.5735 & 0.7495 & 0.6885 & 0.5938 \\
MUDA & 0.4972 & 0.6299 & 0.8268 & 0.7116 & 0.6110 & 0.4522 & 0.5788 & 0.7602 & 0.6898 & 0.5940 \\
\midrule
UKD & 0.5061 & 0.6477 & 0.8301 & 0.7145 & 0.6218 & 0.4610 & 0.5925 & 0.7714 & 0.6962 & 0.6015 \\
UECF & 0.5216 & 0.6699 & 0.8585 & 0.7246 & 0.6313 & 0.4824 & 0.6210 & 0.8055 & 0.7055 & 0.6124 \\
\midrule
SIDA & 0.5143 & 0.6626 & 0.8499 & 0.7210 & 0.6241 & 0.4705 & 0.6105 & 0.7922 & 0.7001 & 0.6045 \\
DAMCAR & 0.5191 & 0.6650 & 0.8498 & 0.7206 & 0.6277 & 0.4788 & 0.6152 & 0.7954 & 0.7022 & 0.6088 \\
\midrule
\textbf{GPL} & \textbf{0.5254} & \textbf{0.6777} & \textbf{0.8680} & \textbf{0.7299} & \textbf{0.6338} & \textbf{0.4912} & \textbf{0.6324} & \textbf{0.8198} & \textbf{0.7112} & \textbf{0.6175} \\
\bottomrule
\end{tabular}}
}
\label{table:main_results}
\end{table*}

\section{Experiments}
\label{gen_inst}

\subsection{Setup}

\subsubsection{\textbf{Datasets}} We evaluate our method on two datasets.

\noindent \textbf{Industrial Dataset}:  
We construct an industrial-scale dataset comprising authentic, high-quality user interaction logs collected from \emph{Taobao}\footnote{\url{https://www.taobao.com}}. 
The dataset spans a continuous 14-day period and includes approximately 200 million users and 20 million items, yielding around 30 billion interaction records. The first 13 days of data are used for training, and the final day is reserved for validation.  
Following the production serving pipeline, the \emph{exposed dataset} consists of items actually displayed to users, while the \emph{unexposed dataset} contains items retrieved during the recall stage but filtered out before display.

\noindent \textbf{Taobao-MM}\footnote{\url{https://huggingface.co/datasets/TaoBao-MM/Taobao-MM}}:  
Taobao-MM~\cite{taobao-mm} is an official large-scale recommendation dataset that incorporates multimodal item representations. In our experiments, we utilize all 8.79 million users and 35.4 million items. User-side features include user ID, age, and gender; item-side features comprise item ID and category. For each user, all items with which they have not interacted are treated as unexposed candidates.



\subsubsection{\textbf{Baselines \& Implementation Details}}
In line with recent studies, we benchmark our method against several representative strategies that exploit unexposed items. Since public datasets lack ranking models, we train a ranking model on exposed dataset to serve as the teacher for those baselines that require pseudo-labels. 
\begin{itemize}[leftmargin=*, itemsep=0pt]
    \item \textbf{Binary Classification (BC)}: A fundamental baseline that trains the model directly on exposed data, treating user clicks as positive labels while ignoring the inherent selection bias.
    \item \textbf{Knowledge Distillation (KD)}: This approach uses the ranking model’s prediction scores as soft pseudo-labels to train the front-end model~\citep{qin2022rankflow}. 
    \item \textbf{Transfer Learning (TL)}: This method fine-tunes the item embedding tower with pseudo-labels on unexposed data while freezing the user tower~\citep{pan2009survey}.
    \item \textbf{Modified Unsupervised Domain Adaptation (MUDA)}: It introduces a threshold-based pseudo-label filtering mechanism that converts the continuous scores of the ranking model into reliable binary labels for training~\citep{wang2023empirical}.
    \item \textbf{Adversarial Domain Adaptation}: This line of work aligns the distributions of exposed and unexposed data by training a domain discriminator with a gradient reversal layer~\citep{goodfellow2014generative}. We instantiate this category with two strong baselines, i.e., \textbf{UKD}~\citep{xu2022ukd} and \textbf{UECF}~\citep{li2025unbiased}.
    \item \textbf{Selective Adversarial Domain Adaptation}: These methods identify informative unexposed items as a target domain via random walks on a user–item bipartite graph and assign robust labels to them through adversarial domain adaptation. We select two SOTA methods \textbf{SIDA}~\citep{wei2024enhancing} and \textbf{DAMCAR}~\citep{lin2024mitigating} as baselines.
\end{itemize}
All models are trained using the Adam optimizer for dense parameters and Adagrad for sparse parameters, with an initial learning rate of $1\times10^{-3}$. Qwen2.5-0.5B~\citep{qwen} is adopted for user interest-anchor generation, and the impact of different LLM backbones is further explored in Section~\ref{Model Configuration Study} and Table~\ref{tab:llm_comp}.

For Industrial Dataset, the input features include user ID, gender, item ID, item category, and the user behavior sequence.  
Item information is encoded using a frozen CLIP model~\cite{radford2021learning}. The RQ-VAE is configured with three quantization layers, each equipped with a codebook of 8,192 entries.

For Taobao-MM Dataset, we adopt SCL-based multimodal embeddings~\cite{sheng2024enhancing} to train the RQ-VAE, which employs a two-layer quantization structure with 200 codebook entries per layer. The input features consist of user ID, age, gender, item ID, item category, and the user behavior sequence.

\subsubsection{\textbf{Metrics}}
In offline experiments, we evaluate top-$K$ recommendation quality using \textbf{Hit Rate (HR@$K$)} with $K \in \{3, 5, 10\}$, where HR@$K$ is calculated on exposed samples for each user. We also assess overall ranking performance via \textbf{AUC} and \textbf{Group AUC (GAUC)}~\citep{DIN}. GAUC computes the average of per-user AUCs to mitigate bias from highly active users and better reflect the typical user experience. Additionally, to measure pseudo-label quality, we generate pseudo-labels for exposed items and compute the AUC between these pseudo-labels and real user feedback, denoted as $\text{\textbf{AUC}}^{*}$.

In online experiments, we measure utility and user engagement.  
Utility metrics—\textbf{Item Page Views (IPV)}, \textbf{Click-Through Rate (CTR)}, and \textbf{Click-Through Conversion Rate (CTCVR)}—capture the system’s effectiveness in driving actions and revenue.  
Engagement is assessed via \textbf{UV3} (users scrolling through >200 items per session) and \textbf{Dwell Time (DT)}, which reflect interaction depth and duration. Higher engagement suggests stronger interest and improved long-term retention.

\subsection{Overall Performance Comparison}
Table~\ref{table:main_results} compares \textbf{GPL} with a range of baselines on our industrial dataset, yielding three key insights.

First, the \textbf{BC} baseline—trained exclusively on exposed interactions—achieves the lowest performance across all metrics. This stark underperformance empirically validates a central premise of our work: unexposed candidates are not merely negative samples but contain rich signals about user preferences. By ignoring them, BC discards potentially relevant items that were never shown due to system limitations, thereby reinforcing exposure bias and limiting recommendation diversity.

Second, pseudo-labeling methods (\textbf{KD}, \textbf{TL}, \textbf{MUDA}) remain suboptimal. Their teacher models, trained only on exposed data, inherit and propagate exposure bias into pseudo-labels, creating a feedback loop that amplifies popularity bias and suppresses discovery of novel interests.

Third, \textbf{GPL} significantly outperforms all baselines, including advanced domain adaptation approaches (\textbf{UKD}, \textbf{UECF}, etc.). Unlike coarse distribution alignment, GPL generates user-specific interest anchors via an LLM and assigns uncertainty-calibrated relevance scores to individual unexposed items. The consistent gains confirm that semantically grounded, dual-label supervision—combining observed clicks with LLM-informed signals—enables more robust and diverse recommendations.

\begin{table}[t]
    \caption{Ablation study of GPL on the Industrial Dataset. }
    \vspace{-8pt}
    \centering
    \scalebox{0.8}{
    \setlength{\tabcolsep}{1.5mm}
    {\begin{tabular}{l|cc|c|c|c}
\toprule
\multicolumn{1}{c|}{\multirow{2}{*}{Method}} & \multicolumn{2}{c|}{HR} & \multirow{2}{*}{AUC} & \multirow{2}{*}{GAUC} & \multirow{2}{*}{$\text{AUC}^*$} \\
\cmidrule{2-3}
 & @3 & @10 & & & \\
 \midrule
w/o Semantic IDs & 0.5158 & 0.8592 & 0.7221 & 0.6252 & 0.6194\\
w/o Multimodal Encoder & 0.5094 & 0.8540& 0.7219 & 0.6236 & 0.6167\\
w/o Max-Pooling & 0.5072 & 0.8537& 0.7184 & 0.6220 & 0.6023\\
\midrule
w/o Confidence & 0.5215& 0.8639 & 0.7223 & 0.6279 & -\\
w/o Semantic Dispersion & 0.5233& 0.8656 & 0.7269 & 0.6294 & -\\
w/o Historical Consistency & 0.5244& 0.8667 & 0.7281 & 0.6317 & -\\
w/o LLM-Intrinsic Confidence & 0.5237& 0.8662 & 0.7275 & 0.6304 & -\\
\midrule
w/o Actual Labels & 0.3687 & 0.6933 & 0.6239 & 0.5863 & -\\
\midrule
\textbf{GPL} & \textbf{0.5254} & \textbf{0.8680} & \textbf{0.7299} & \textbf{0.6338} & \textbf{0.6275}\\ 
\bottomrule
\end{tabular}}
    }
\label{tab:Ablation}
\vspace{-10pt}
\end{table}

\subsection{Ablation Study}
Table~\ref{tab:Ablation} shows that every component of GPL is essential: removing any module leads to consistent performance drops.

\begin{itemize}[leftmargin=*, itemsep=0em]
    \item \textbf{w/o Semantic IDs}: The RQ-VAE tokenization is removed. The LLM is trained to directly predict the next raw item ID from historical item ID sequences.

    \item \textbf{w/o Multimodal Encoder}: The frozen multimodal encoder $f_e$ is replaced with standard learnable ID embeddings from the item embedding layer.
    
    \item \textbf{w/o Max-Pooling}: Max-pooling over anchor similarities is replaced with mean-pooling. Pseudo-labels are computed as the average (rather than maximum) cosine similarity between an unexposed item and all interest anchors.
    
    \item \textbf{w/o Confidence}: The confidence-aware weighting is disabled. All pseudo-labels are assigned uniform weight $w_{u,h} = 1$ in $\mathcal{L}_{\text{pl}}$, assuming equal reliability.
    
    \item \textbf{w/o Semantic Dispersion}: The semantic dispersion term ($\sigma_u$) is excluded from the confidence weight computation.
    
    \item \textbf{w/o Historical Consistency}: The historical consistency term ($\rho$) is removed from the confidence weight computation.
    
    \item \textbf{w/o LLM-Intrinsic Confidence}: The LLM-intrinsic confidence term ($\text{conf}$) is dropped from the confidence weight computation.
    
    \item \textbf{w/o Actual Labels}: The actual-label loss $\mathcal{L}_{\text{al}}$ (Eq~(\ref{eq:objective})) is removed. The model is trained solely on pseudo-labels.
\end{itemize}
\vspace{+3pt}
\noindent\textbf{Effect of pseudo label quality}.
In Table~\ref{tab:Ablation}, the first three variants (rows 1–3) assess the contribution of each component in pseudo-label generation. Notably, all components are essential: removing any one results in a significant drop in $\text{AUC}^{*}$, indicating degraded pseudo-label quality, which in turn leads to reduced performance of the final ranking model. Among these, the \textit{w/o Max-Pooling} variant—replacing max-pooling with mean-pooling—causes the largest degradation, reducing HR@3 by 0.182. This supports the intuition that focusing on the single most relevant interest anchor yields more discriminative signals than averaging across all anchors.

\vspace{+3pt}
\noindent\textbf{Impact of the Confidence}. Rows 4–7 in Table~\ref{tab:Ablation} evaluate the role of confidence weighting, which does not affect pseudo-label quality directly but is crucial during joint optimization. Specifically, the \textit{w/o Confidence} variant reduces HR@3 from 0.5254 to 0.5215, confirming that calibrating the LLM’s raw outputs is essential for effective pseudo-labeling. We further dissect the confidence weighting mechanism into its three uncertainty components. Among them, removing Semantic Dispersion causes the largest performance drop, indicating that LLM-generated interest anchors exhibit substantial user-wise variability in semantic coherence. Explicitly modeling this dispersion is therefore critical for robust uncertainty estimation.

\vspace{+3pt}
\noindent\textbf{Importance of Actual Label}.
The most severe degradation occurs in the \textit{w/o Actual Labels} setting, where HR@3 plummets from 0.5254 to 0.3687. This dramatic drop underscores that supervision from observed interactions is indispensable—not only for providing ground-truth signals but also for stabilizing training and preventing model drift in the presence of noisy pseudo-labels.

\subsection{Model Configuration Study}
\label{Model Configuration Study}
In this section, we evaluate the impact of different LLM backbones and tokenization strategies to justify the cost-effectiveness and architectural choices of GPL.

\vspace{5pt}
\noindent{\textbf{Effect of LLM Model Size}}.
We compare three Qwen2.5 variants~\citep{qwen} for interest anchor generation: 0.5B, 1.8B, and 7B. As shown in Table~\ref{tab:llm_comp}, larger models improve both the final AUC on real user interactions and the pseudo-label quality, measured by $\text{AUC}^{*}$, which is the AUC between pseudo-labels and real feedback on exposed items. The 7B model achieves the highest AUC of 73.54 and $\text{AUC}^{*}$ of 63.33 but requires 5.25\,Gflops/token. In contrast, the 0.5B model attains a competitive AUC of 72.99 with only 0.30\,Gflops/token and $\text{AUC}^{*} = 62.57$. \textbf{Although $\text{AUC}^{*}$ is lower than AUC, indicating imperfect pseudo-labels, it remains sufficiently aligned with user feedback to effectively guide training}. Given Taobao's massive traffic scale, the 0.5B variant offers the best trade-off between performance and inference efficiency.

\begin{table}[h]
    \caption{Performance comparison across LLM backbones.}
    \vspace{-8pt}
    \centering
    \scalebox{1.0}{
    \begin{tabular}{lccc}
        \toprule
        LLM-backbone & Gflops/token & AUC & $\text{AUC}^{*}$ \\
        \midrule
        Qwen2.5-0.5B & 0.30 & 0.7299 & 0.6257 \\
        Qwen2.5-1.8B & 0.98 & 0.7317 & 0.6286 \\
        Qwen2.5-7B & 5.25 & 0.7354 & 0.6333 \\
        \bottomrule
    \end{tabular}}
    \label{tab:llm_comp}
    \vspace{-10pt}
\end{table}

\begin{figure}[t]
\vspace{-6pt}
\begin{center}
\includegraphics[width=0.95\columnwidth]{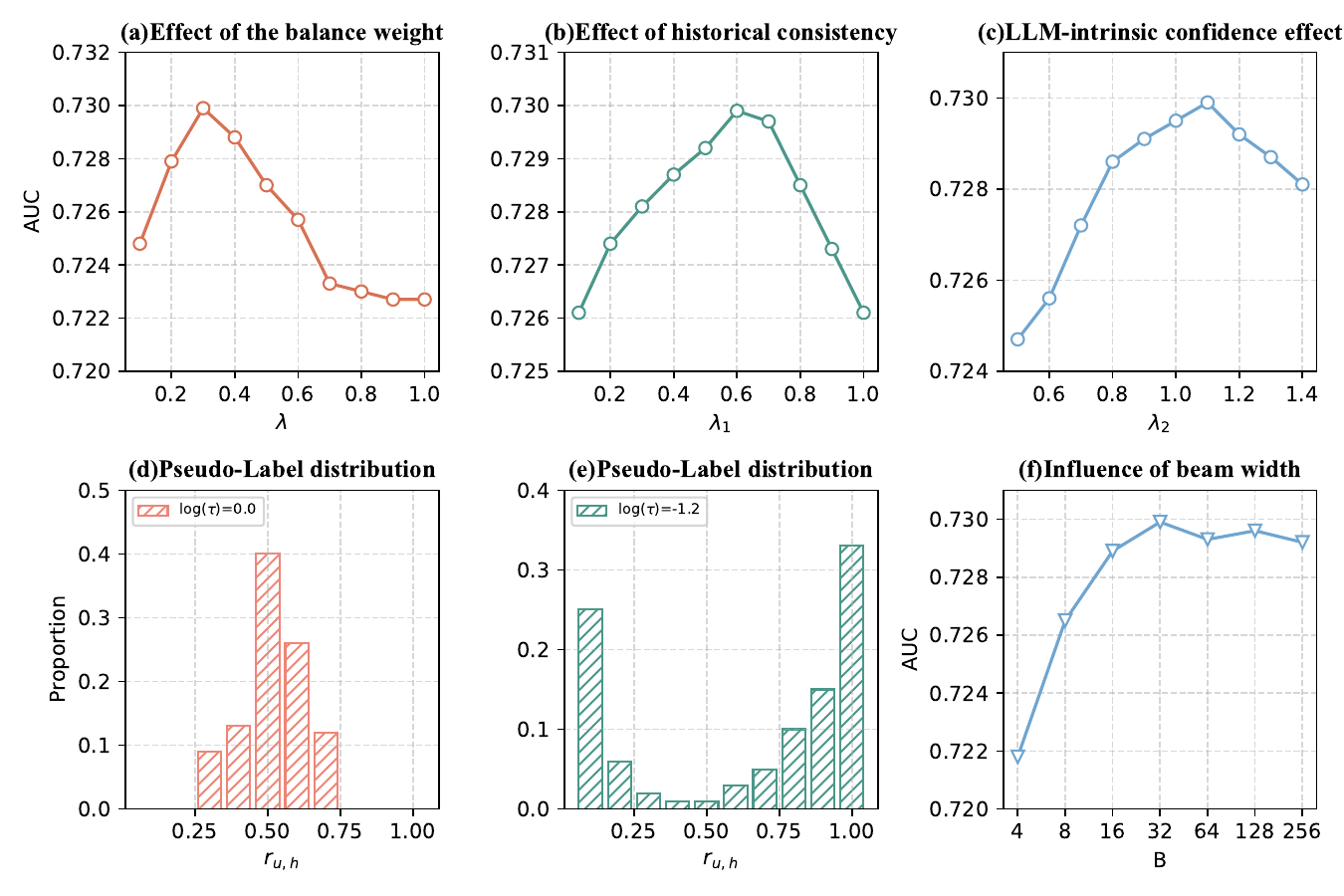}
\vspace{-6pt}
\end{center}
\caption{We conduct a parameter sensitivity analysis of the balance weight~$\boldsymbol{\lambda}$, the historical consistency weight~$\lambda_1$, and the LLM-intrinsic confidence weight $\lambda_2$. The effects of the beam size~$\boldsymbol{B}$ and temperature~$\tau$ are also investigated. AUC performance is plotted for all parameters across their respective ranges.}
\label{fig:Parameter}
\vspace{-8pt}
\end{figure}

\vspace{5pt}
\noindent\textbf{Sensitivity to the balance weight \boldmath$\lambda$ in joint optimization.} As shown in the Figure~\ref{fig:Parameter}-(a), varying $\lambda$ exhibits a clear unimodal trend. Small $\lambda$ values collapse training into exposure-only supervision, reducing interest coverage and lowering AUC and HR@3, whereas large values over-trust generated labels, amplifying noise and distribution shift, again harming performance. The optimum occurs at moderate $\lambda$ (0.1–1.0 in our setting), where actual and estimated supervision are balanced. Without confidence weighting, the optimum shifts lower and the right tail drops more steeply, showing that it mitigates noise from unreliable anchors, broadens $\lambda$'s stable range, and raises the performance ceiling.

\vspace{+5pt}
\noindent\textbf{Analysis of $\lambda_1$, $\lambda_2$, and temperature $\tau$}. As shown in Figure~\ref{fig:Parameter}-(b), AUC peaks around $\lambda_1 \in [0.4, 0.6]$ and declines beyond 0.6, since over-emphasizing historical similarity limits anchor diversity and exacerbates SSB. Peak AUC for $\lambda_2$ is achieved at 1.1 (Fig.~\ref{fig:Parameter}-(c)); larger values amplify hallucinations. For temperature $\tau$, lower values sharpen the label distribution but saturate the sigmoid and discard intra-class information; peak AUC occurs at $\log(\tau) \approx -1.2$ (Fig.~\ref{fig:Parameter}-(d)(e)).

\begin{table}[t]
\vspace{-4pt}
    \caption{Online A/B Test Performance Gains.}
    \vspace{-8pt}
    \centering
    \scalebox{0.9}{
    \setlength{\tabcolsep}{2mm}
    {\begin{tabular}{c|ccccc}
\toprule
Method & IPV & CTR & CTCVR & UV3 & DT \\
\midrule
GPL & +3.53\% & +3.07\% & +2.51\% & +1.50\% & +2.16\%  \\
 \bottomrule
\end{tabular}}
}
\label{tab:online_results}
\vspace{-10pt}
\end{table}

\vspace{5pt}
\noindent\textbf{Effect of beam size $B$.} As shown in Fig.~\ref{fig:Parameter}-(f), increasing $B$ steadily boosts metrics until saturation at $B=32$ (our default), beyond which extra anchors mostly contribute near-duplicates. Without confidence weighting, performance peaks earlier and degrades faster, indicating that larger anchor sets require uncertainty-aware moderation to remain beneficial.

\subsection{Online Experiments.}
\subsubsection{\textbf{Overall Effectiveness in Online Deployment}}
To rigorously assess GPL’s real-world effectiveness and scalability, we deployed it in the pre-ranking stage of Taobao’s “Guess What You Like” homepage feed and conducted a two-week online A/B test with hundreds of millions of daily active users. As shown in Table~\ref{tab:online_results}, GPL achieves consistent gains across all metrics: CTR (+3.07\%) and IPV (+3.53\%) improve significantly, confirming more engaging recommendations. The +2.51\% lift in CTCVR indicates these clicks are high-quality, driving more purchases and business value. Gains in UV3 (+1.50\%) and Dwell Time (+2.16\%) further suggest users find the content more compelling, leading to deeper exploration and longer sessions.

\subsubsection{\textbf{Impact on SSB and Generalization}} \label{ssb_dis}
\begin{figure}[t]
\vspace{-4pt}
\centering
\includegraphics[width=1\columnwidth]{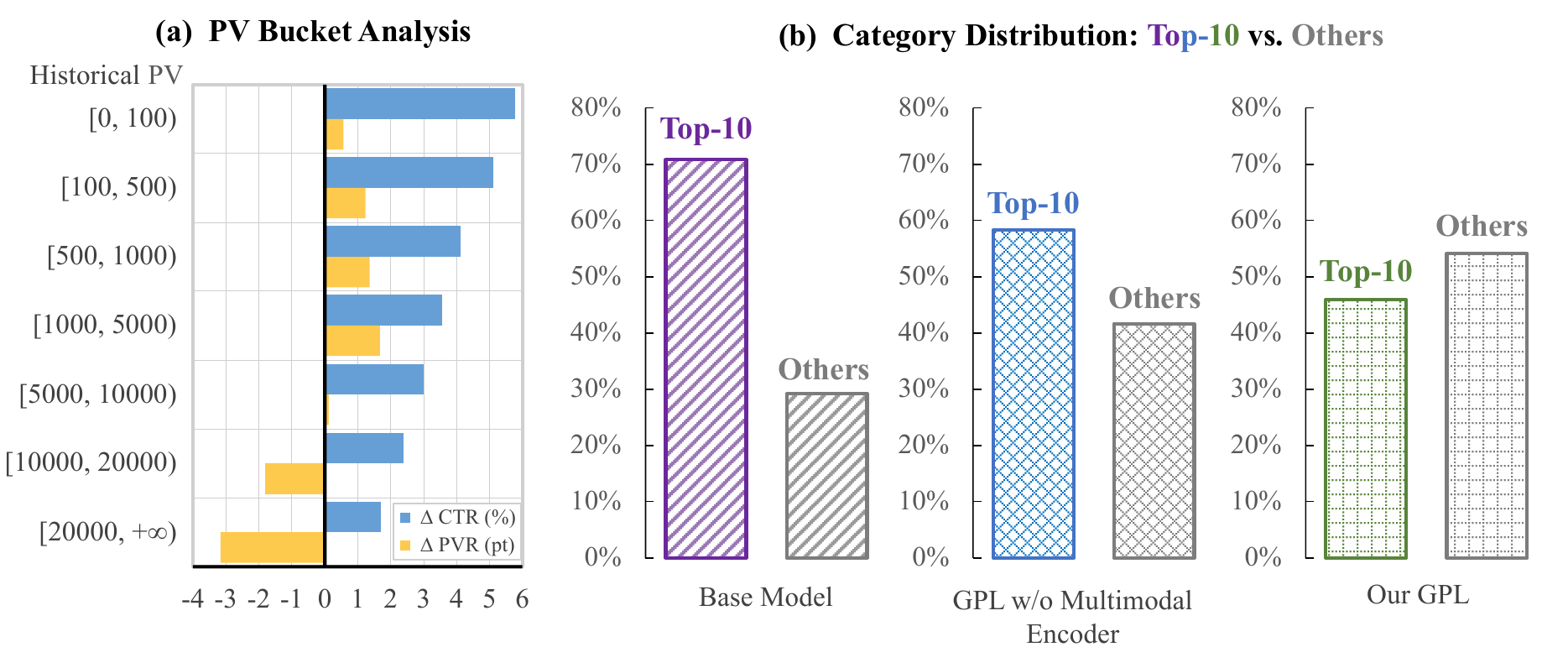}
\vspace{-14pt}
\caption{Fine-Grained Online A/B Analysis of GPL vs. Baseline.
(a) Relative CTR lift (\%) and absolute PVR change (percentage points, pt) across historical page-view (PV) buckets.
(b) Category distribution of exposed items.}
\label{fig:online_analysis}
\vspace{-8pt}
\end{figure}

To validate our hypothesis that GPL mitigates selection bias and enhances discovery, we conduct a fine-grained online analysis (Figure~\ref{fig:online_analysis}).

First, we partition items into seven buckets by historical Page Views (PV). Figure~\ref{fig:online_analysis}(a) shows that GPL achieves increasing CTR lift for lower-PV (long-tail) items, indicating its ability to surface relevant but previously overlooked content. The pseudo-labels provide crucial supervision to counteract popularity bias.

Second, we examine category diversity. The baseline concentrates 70.80\% of recommendations in the top-10 categories, leaving only 29.20\% for the remaining 100+ categories. In contrast, GPL reduces the dominant share to 45.91\%—a >35\% relative drop. The w/o Multimodal Encoder variant lies in between, suggesting two complementary effects: (1) LLM-based interest generation reduces reliance on exposure history and expands coverage; (2) alignment with a content-rich multimodal space further debiases representation, enabling broader cross-category exploration.


\section{Conclusion}
In this paper, we presented GPL, a framework that jointly leverages actual labels from exposed interactions and pseudo labels for unexposed candidates generated via an LLM-driven preference estimation module. 
By converting simulated feedback into explicit supervision, GPL alleviates exposure bias, improves data efficiency, and models a broader spectrum of user interests. 
Deployed on a large-scale industrial platform, our approach achieved a $3.07\%$ CTR lift in online A/B testing, demonstrating its practical value and scalability.
\bibliographystyle{ACM-Reference-Format}
\bibliography{sample-base}

\end{document}